\documentclass[a4paper,twocolumn]{jpsj3}
\usepackage{txfonts,amsmath}
       
\title{Theory of Antiferromagnetic Order in High-$T_c$ Oxides:\\
An Approach Based on Ginzburg-Landau Expansion}

\author{Masahiko Hayashi$^1$\thanks{E-mail: m-hayashi@ed.akita-u.ac.jp}, 
Yasunari Tanuma$^2$ and 
Kazuhiro Kuboki$^3$}
\inst{$^1$Faculty of Education and Human Studies, 
Akita University, Akita 010-8502, Japan \\
$^2${Faculty of Engineering and Resource Science, Akita University, Akita 010-8502, Japan}\\
$^3${Department of Physics, Kobe University, Kobe 657-8501, Japan}} %\\

\abst{The mean-field phase diagram of antiferromagnetic order in 
$t$-$J$ model has been examined, using the free energy obtained 
by Ginzburg-Landau (GL) expansion. 
We extended the usual GL theory in 
two ways: 
First, we have included higher order terms with respect to the 
spatial derivative (or wave number) 
to incorporate the incommensurate 
antiferromagnetic order. 
Second, we have also included 
higher order terms with respect to the order parameter amplitude, 
in order to treat the first order phase transition 
between paramagnetic and antiferromagnetic phase, 
which appears at some 
doping rates. 
We found the possibility of tricritical point and 
critical endpoint in the magnetic phase diagram of the high-$T_c$ oxides 
associated with the commensurate and incommensurate antiferromagnetic order. 
The possible effects of thermal fluctuations and randomness (spin glass) 
are also discussed qualitatively based on the GL free energy. }

\kword{high-$T_c$ oxides, GL theory, $t$-$J$ model, antiferromagnetism, phase diagram}

\begin{document}
\maketitle

\section{Introduction}

From the early stage of the research of high-$T_c$ oxides, 
the magnetic structure of these materials has been considered as a 
key factor in understanding the mechanism of superconductivity. 
Until now, various intriguing results, both theoretical and 
experimental, have been obtained and the research is still going on. 
One of the most interesting phenomena may be the pseudogap, 
which has not been completely 
understood until now 
(see Refs. \citen{Timusk:1999wp,Norman:2005vr} and references therein). 

Among the various theoretical efforts to understand the 
mechanism of high-$T_c$ oxides, 
the $t$-$J$ model is one of the promising model to 
capture the natures of strong correlations 
\cite{Anderson:2007ta,Zou:1988ul,Suzumura:1988wd,Nagaosa:1990ut,Lee:1992wg}. 
The magnetic responses in NMR or neutron scattering experiments have 
been analyzed based on $t$-$J$ model 
\cite{Tanamoto:1991tz, Tanamoto:1993tz, Tanamoto:1994ww}. 
The antiferromagnetic order has been also studied\cite{Inaba:1996fm}. 

Recently, the synthesis of multilayered cuprates has 
revealed that the antiferromagnetic phase is more robust 
in multilayered materials \cite{JULIEN:2003ec,Kitaoka:2011kj,Mukuda:2011km} 
than single-layered ones \cite{Sanna:2004vi,Iyo:2007wr}, 
thus suggesting the importance of dimensionality on 
the pseudogap and antiferromagnetism in high-$T_c$ oxides 
(see Ref. \citen{Mukuda:2011km} and references therein). 
Several authors have studied the effects of interlayer coupling to 
understand these features from theoretical point of view
\cite{Chakravarty:2004um,Zaleski:2005wl,Mori:2005vn,Mori:2006ul}. 
Another important aspect of the multilayered systems is 
the coexistence of superconductivity and antiferromagnetism in 
a certain region of the phase diagram 
\cite{Mukuda:2008uo,Shimizu:2009ue,Shimizu:2009wj}. 
Such phenomenon has been studied theoretically also using $t$-$J$ model 
\cite{Himeda:1999ux,Yamase:2004us,Yamase:2011kf,Kuboki:2013ho}.

The researches to refine understanding of the $t$-$J$ model are also 
currently in progress. 
Recently, Yamase {\it et al.} have studied 
antiferromagnetic order in $t$-$J$ model 
incorporating the possibility of the incommensurate order\cite{Yamase:2011kf}.  
It has been shown that the 
incommensurate antiferromagnetic (IC-AF) order can 
be stable in a certain range of doping 
in addition to the ordinary commensurate antiferromagnetic (C-AF) order 
which was first studied in Ref. \citen{Inaba:1996fm}. 

In this paper, we extend the study of Ref. \citen{Yamase:2011kf} and 
re-examine the details of the magnetic phase diagram of $t$-$J$ model. 
Following the study by one of the present authors (KK) 
\cite{Kuboki:2013ho}, we derive the Ginzburg-Landau (GL) free energy of 
antiferromagnetic order based on the $t$-$J$ model. 
We extend the GL free energy in the following two ways:  
Firstly, we introduce the higher order terms of 
gradient expansion of the order parameter. 
In the usual GL free energy, 
the spatial derivative (or the wave number $\vec{q}$ dependence) 
is kept only to the second order. 
However, the IC-AF order is the \lq\lq finite-$\vec{q}$ 
antiferromagnetic order\rq\rq and we need to 
keep higher order terms to treat it. 
Therefore, we keep up to the fourth order in $\vec{q}$. 
Secondly, the expansion with respect to the order parameter amplitude 
is also extended. 
Actually, we will find that 
the free energy as a function of the 
order parameter amplitude 
is strongly nonlinear and, in some case, 
the first order transition can occur. 
To incorporate these features, we include 
up to the sixth order in the amplitude, 
instead of the ordinary fourth-order expansion. 

In order to concentrate on the 
antiferromagnetic order, we neglect the other orders 
such as RVB order. 
This treatment, although limited in quantitative accuracy, 
may be sufficient to capture the 
overall structure of the magnetic phase diagram of the high-$T_c$ oxides. 
For example, the antiferromagnetic phase 
transition lines obtained in this paper 
(Fig. \ref{phase_diag} of this paper) qualitatively coincides 
with those obtained by Inaba {\it et al.} \cite{Inaba:1996fm} 
and Yamase {\it et al.}\cite{Yamase:2011kf}. 
also taking the RVB order into account
 (see especially Fig. 2 of Ref. \citen{Yamase:2011kf}). 

This paper is organized as follows: 
In Sect. 2, we describe the model and the method to 
derive GL free energy. 
In Sect. 3, basic feature of our nonlinear GL free energy is discussed. 
In Sect. 4, the magnetic phase diagram of the high-$T_c$ oxides is discussed 
based on the derived GL free energy. 
In Sect. 5, we summarize our results. 
Some calculations are given in the Appendices. 

We adopt units $k_B=\hbar=1$ throughout. 

\section{Derivation of Ginzburg-Landau Free Energy}

Here we derive the Ginzburg-Landau (GL) free energy which describes the 
antiferromagnetic order in high-$T_c$ oxides. 

\subsection{Model}

We start from the Hamiltonian of the $t$-$J$ model, 
\begin{align}
H &= -t \sum_{j,\sigma} \sum_{\hat\delta = \pm x, \pm y} \tilde{c}^\dagger_{j+\hat\delta\sigma}
 \tilde{c}_{j\sigma}+ \frac{J}{2} \sum_{j}\sum_{\hat\delta=\pm \hat{x},\pm \hat{y}}
\vec{S}_{j}\cdot\vec{S}_{j+\hat\delta}
\end{align}
where $t$ is the hopping integral, $J$ is the superexchange interaction and 
the double occupancy is assumed to be removed from the Fock space. 
We consider a simple square lattice, whose lattice sites are numbered by  
$j$, and $j+\hat\delta$ correspond to the nearest-neighbor sites of $j$. 
(In this paper, we consider the case of nearest-neighbor hopping only. )
Operator $\tilde{c}_{j\sigma}^\dagger$ ($\tilde{c}_{j\sigma}$) is 
the creation (annihilation) operator of an electron on the $j$-th site 
with spin $\sigma$ ($\sigma=\uparrow$ or $\downarrow$). 
Here we adopt the slave-boson method, 
in which an electron operator is decomposed as $\tilde{c}_{j\sigma}= b^\dagger_j 
f_{j\sigma}$, 
where $b^\dagger_j$ is the creation operator of a holon (boson) 
and $f_{j\sigma}$ in the annihilation operator of a spinon (fermion).  
The sum of holon number and spinon number on each site should be 
unity so that the double occupancy is absent. 
The spin operator is then given by 
$\vec{S}_j = (1/2)(f_{j\alpha}^\dagger \vec{\sigma}_{\alpha\beta}f_{j\beta})$, 
where $\alpha,\beta=\{\uparrow,\downarrow\}$ and 
$\vec{\sigma}_{\alpha\beta}$ are the Pauli matrices. 

Next we introduce mean-field approximation\cite{Suzumura:1988wd}. 
Since we consider the region where the doping rate $\delta$ is 
sufficiently high ($\delta > 0.14$), 
we assume that holons bose-condense at a higher temperature 
than the antiferromagnetic transition temperature, and 
put  $\langle b_j^\dagger b_{k}\rangle =\langle b_j\rangle^2 =\delta$ 
in the following. 
Here $\langle \cdots \rangle$ denotes statistical average. 
We also introduce the bond order parameter 
$\chi=2 \langle f_{j\uparrow}^\dagger f_{j+\hat\delta\uparrow}\rangle=
2 \langle f_{j\downarrow}^\dagger f_{j+\hat\delta\downarrow}\rangle$, 
which is assumed to be spatially constant. 
Then the microscopic Hamiltonian for the \lq\lq spinon\rq\rq part, 
after decoupling the superexchange term, 
becomes $H=H_0+H_I$, 
\begin{align}
H_0 =& - \sum_{j,\sigma} \sum_{\hat\delta = \pm x, \pm y}
\tilde{t}
f_{j+\hat\delta\sigma}^\dagger f_{j\sigma}
-\mu\sum_{j\sigma} 
f_{j\sigma}^\dagger f_{j\sigma}
\nonumber\\
H_I =& J\sum_{j}\sum_{\hat\delta=\pm \hat{x},\pm \hat{y}}
\left\langle\vec{S}_{j}\right\rangle\cdot\vec{S}_{j+\hat\delta}
-\frac{J}{2}\sum_{j}\sum_{\hat\delta=\pm \hat{x},\pm \hat{y}}
\left\langle\vec{S}_{j}\right\rangle\cdot\left\langle\vec{S}_{j+\hat\delta}\right\rangle
%\nonumber\\
%H_{\rm ext} =& \mu \sum_j \vec{H}_j \cdot\vec{S}_j
\label{Hamiltonian}
\end{align}
where $\mu$ is the chemical potential of spinons. 
The parameter $\tilde{t}$ is given by $\tilde{t}=t \delta +(3/8)J\chi$, 
where $\chi$ and $\mu$ are determined self-consistently from 
\begin{align}
\chi = \frac{1}{N}\sum_{\vec{k}}
\gamma_{\vec{k}} f(\xi_{\vec{k}}),\,\,\,
\delta = 1 - \frac{2}{N}\sum_{\vec{k}} f(\xi_{\vec{k}}).
\end{align}
Here $N$ is the total number of the lattice sites, 
$\gamma_{\vec{k}} = \cos k_x +\cos k_y$, 
$\xi_{\vec{k}} = - 2\tilde{t}\gamma_{\vec{k}}-\mu$ 
 and $f(z) = (1+e^{z/T})^{-1}$ with $T$ being the temperature. 
The Fourier transform is defined as 
$f_{\vec{k}} = \sqrt{N}^{-1}\sum_{j} f_{j}e^{-i \vec{k}\cdot\vec{r}_j}$, 
where $\vec{r}_j$ is the coordinate of 
the $j$-th lattice site and 
$\vec{k}$-sum is taken over the first Brillouin zone defined by 
$|k_x|, |k_y|<\pi$. 
In this paper we take the lattice spacing to be unity.

We introduce the order parameter $m_j$ by 
$\langle \vec{S}_j \rangle \equiv m_j\,\vec{n}_j$, 
where $\vec{n}_j$ is a unit vector. 
We assume that the spatial variation of $\vec{n}_j$ is 
so gradual that we can treat it as an adiabatic change for spinons. 
We take the quantization axis ($z$-axis) of the spins to be parallel to $\vec{n}_j$, 
and $m_j$ can be written as  
$\langle f_{j\uparrow}^\dagger f_{j\uparrow}-f_{j\downarrow}^\dagger f_{j\downarrow}
\rangle/2$. 
Within this approximation, 
$H_I$ is given as 
\begin{align}
H_I&=\frac{J}{2}\sum_{j}\sum_{\hat\delta=\pm \hat{x},\pm \hat{y}}
m_{j+\hat\delta}(f_{j\uparrow}^\dagger f_{j\uparrow}-f_{j\downarrow}^\dagger f_{j\downarrow})
\nonumber\\
&-\frac{J}{2}\sum_{j}\sum_{\hat\delta=\pm \hat{x},\pm \hat{y}} m_j m_{j+\hat\delta}.
\label{MFH}
\end{align}
Later, we will recover the vector degree of freedom of the spins at the GL level. 

Throughout this paper, we take $J$ to be the unit of energy and 
put $t=4 J$.

\subsection{Perturbative Expansion}

Using path-integral framework, 
the free energy is obtained as follows: 
We write the partition function as, 
\begin{align}
Z &=  \int{\cal D}[f_{\vec{k}\omega_n\sigma}^*,f_{\vec{k}\omega_n\sigma}]
e^{-S-S_0}
\nonumber\\
S &= -\sum_{\omega_n,\vec{k},\vec{k}',\sigma}
\left[G_{\vec{k},\omega_n}^{-1} \delta_{\vec{k},\vec{k}'}-\frac{(-1)^\sigma}{\sqrt{N}} J \gamma_{\vec{k}-\vec{k}'}
m_{\vec{k}-\vec{k}'}\right]f_{\vec{k}\omega_n\sigma}^*f_{\vec{k}'\omega_n\sigma}
\nonumber\\
&= -\sum_{\omega_n,\sigma} { }^t \vec{f}_{\omega_n\sigma}^*\cdot
\left[\hat{G}^{-1}+\frac{(-1)^\sigma}{\sqrt{N}}\hat{\Gamma}\right]\cdot\vec{f}_{\omega_n\sigma}
\end{align}
where $G_{\vec{k},\omega_n} = (i\omega_n - \xi_{\vec{k}})^{-1}$ 
with $\omega_n = (2n+1)\pi T$ being the Matsubara frequency, and 
$(-1)^\sigma=1$ $(-1)$ for $\sigma=\uparrow$ ($\downarrow$). 
The constant $S_0$ is given by $-T^{-1}\sum_{\vec{q}}J \gamma_{\vec{q}} |m_{\vec{q}}|^2$. 
In the last line, we used the matrix representation for indices $\vec{k}$ and $\vec{k}'$, 
where 
\begin{align}
\left\{\hat{G}\right\}_{\vec{k},\vec{k}'}
=G_{\vec{k},\omega_n} \delta_{\vec{k},\vec{k}'},\,\,\,
\left\{\hat{\Gamma}\right\}_{\vec{k},\vec{k}'}
=-J \gamma_{\vec{k}-\vec{k}'}
m_{\vec{k}-\vec{k}'}.
\end{align}
The free energy is, then, expressed as
\begin{align}
F&=-T \sum_{\omega_n,\sigma}\left\{{\rm Tr} \ln \left(\hat{G}^{-1} +\frac{(-1)^\sigma}{\sqrt{N}}
\hat{\Gamma}\right)\right\}+TS_0
\nonumber\\
&=-T \sum_{\omega_n,\sigma}{\rm Tr} \left\{\ln \hat{G}^{-1} + \sum_{l=1}^\infty
\frac{(-1)^{l-1}}{l}\left(\frac{(-1)^\sigma\hat{G}\,\hat{\Gamma}}{\sqrt{N}}\right)^l\right\}+TS_0.
\end{align}
From this equation the second order term in $m_{\vec{q}}$ is given by  
\begin{align}
F^{(2)}&=\sum_{\vec{q}}\left(-J \gamma_{\vec{q}} + J^2 T
\gamma_{\vec{q}}^2 \sum_{\vec{k},\omega_n}
G_{\vec{k},\omega_n}G_{\vec{k}-\vec{q},\omega_n}\right)|m_{\vec{q}}|^2
\label{f20}
\\
&=\sum_{\vec{q}}\,\,J\,\gamma_{\vec{q}}\,(1-J\,\gamma_{\vec{q}}\,\chi^{(0)}_{\vec{q}})
|M_{\vec{q}}|^2.
\label{f2}
\end{align}
Here we have substituted $\vec{q}\rightarrow \vec{q}+\vec{Q}$ where 
$\vec{Q}=(\pi,\pi)$ is the reciprocal lattice vector 
(Note the relation $\gamma_{\vec{q}+\vec{Q}}=-\gamma_{\vec{q}}$. ),  
and introduced the staggered (antiferromagnetic) order parameter 
$M_j \equiv m_j e^{i\vec{Q}\cdot\vec{r}_j}$ 
(or $M_{\vec{q}} \equiv m_{\vec{q}+\vec{Q}}$). 
The bare staggered susceptibility $\chi^{(0)}_{\vec{q}}$ is given by 
\begin{align}
\chi^{(0)}_{\vec{q}} &= -  \frac{T}{N}\sum_{\vec{k},\omega_n}
G_{\vec{k},\omega_n}G_{\vec{k}-\vec{q}-\vec{Q},\omega_n}
= -\frac{1}{N}\sum_{\vec{k}}\,\,\frac{f(\xi_{\vec{k}}) - f(\xi_{\vec{k}-\vec{Q}-\vec{q}})}
{\xi_{\vec{k}} - \xi_{\vec{k}-\vec{Q}-\vec{q}}}.
\end{align}
The free energy $F^{(2)}$ is expanded with respect to $\vec{q}$ up to the fourth order. 
From the symmetry argument, the terms appearing in this expansion are proportional to 
$1$, $|\vec{q}|^2$, $q_x^4+q_y^4$, or $q_x^2 q_y^2$.

The higher order terms with respect to the amplitude $|M_j|$ can be derived from the 
following free energy $U$, which is obtained by 
assuming that the order parameter is constant, $M_j = M$ (see Appendix \ref{GLcoeff2}), 
\begin{align}
&U(M) = 2 J M^2
\nonumber\\
&- \frac{2 T}{N}\sum_{\vec{k}\in {\rm MBZ}} 
\left[\ln \left(2 \cosh \frac{E_{\vec{k}}-\mu}{2T}\right) + \ln \left(2 \cosh \frac{E_{\vec{k}}+\mu}{2T}\right)
\right]
\label{potential-eqn}
\end{align}
where $E_{\vec{k}}=\sqrt{(2 \tilde{t}\gamma_{\vec{k}})^2+(2JM)^2}$.  
In this case, the range of $\vec{k}$-sum is the magnetic Brillouin zone (MBZ), 
defined by $|k_x|+|k_y|\leq\pi$. 

The total free energy is expressed as $F = F^{(2)} + \sum_j \bar{U}(M_j)$, 
where $\bar{U}(M)=\beta_4 M^4 + \beta_6 M^6+\cdots$ is obtained by removing 
the second order term from the series expansion of $U(M)$ with respect to $M$, 
since the second order term is already included in $F^{(2)}$. 
Here the constant part independent of $M$ is suppressed. 

\section{Extension of the GL Expansion}

We assume the form of our GL free energy to be
\begin{align}
F = \sum_{\vec{q}}K(\vec{q}) |M_{\vec{q}}|^2 
+\sum_j \left\{\beta_4 M_j^4 + \beta_6 M_j^6\right\}, 
\label{GLF}
\end{align}
where the wave vector dependence of $K(\vec{q})$ is 
given by 
\begin{align}
K(\vec{q}) 
&= A + B |\vec{q}|^2 + C_1 (q_x^4+q_y^4) + C_2 q_x^2 q_y^2. 
\label{k-def}
\end{align}
The coefficients $\beta_4$ and $\beta_6$ are defined in the 
previous section, and  
the calculation of $A$, $B$, $C_1$ and $C_2$ is given 
in Appendix \ref{GLcoeff1}. 
Note that all the coefficients depend on 
$T$ and $\delta$. 

\subsection{GL coefficients}

\begin{figure}[h]
\begin{centering}
\includegraphics[clip,width=8cm]{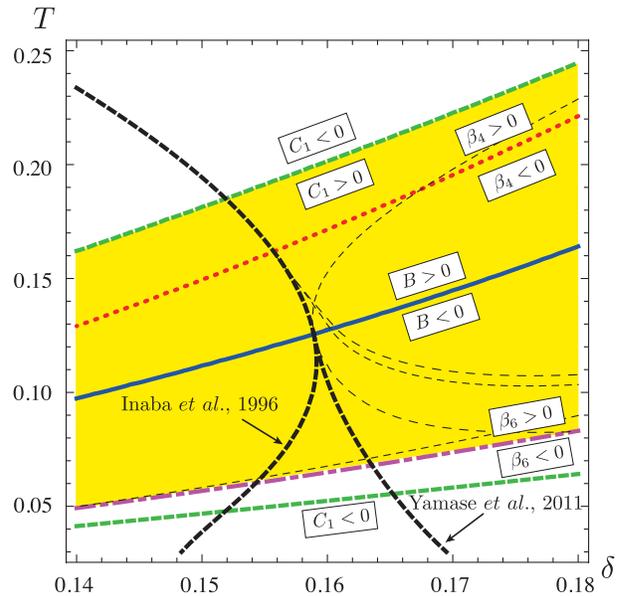} 
\par\end{centering}
\caption{(Color online) Map of the GL coefficients: sign changes of GL coefficients 
are indicated in the figure. 
Bold dashed lines correspond to previously obtained results, 
by Inaba {\it et al.}\cite{Inaba:1996fm} and Yamase {\it et al.} \cite{Yamase:2011kf}
Thin broken lines are shown to clarify the correspondence with Fig. \ref{phase_diag}. 
The painted area (in yellow) is the region where the present GL expansion is well-defined. }
\label{GL_param} %
\end{figure}

Here we discuss the behavior of GL coefficients. 
In this paper we concentrate on the region 
($0.14 < \delta < 0.18$ and $0.03< T < 0.25$) depicted in 
Fig. \ref{GL_param} (color online), 
since the IC-AF order is most significant in this region. 
Unfortunately, whole of this region can not 
be covered by the present treatment. 
The reason is as follows: 
\begin{enumerate}
\item[i)] In order for the $F$ in Eq. (\ref{GLF}) to be stable 
as a function of the amplitude $M_j$, 
at least the highest order term in $M_j$ should be positive. 
In the present case, $\beta_6>0$ is required. 
However, this is not fulfilled in the whole region;  
the boundary between the regions of $\beta_6>0$ 
and $\beta_6<0$ is shown by dashed-dotted line in Fig. \ref{GL_param}. 
\item[ii)] As for the $\vec{q}$-dependence in $K(\vec{q})$, 
the highest order term should be also positive. 
Otherwise, the higher order terms, {\it e.g.}$q_x^6+q_y^6$, may not be negligible. 
This condition is given by 
$C_1>0$ for $C_2 - 2 C_1>0$ and 
$C_1+C_2>0$ for $C_2 - 2 C_1<0$ 
(see the next section for details). 
In Fig. \ref{GL_param}, 
we have depicted the boundary between 
$C_1>0$ and $C_1<0$ (dashed line). 
The value $C_1+C_2$ is always positive in the area of $C_1>0$\cite{Kq}. 
\end{enumerate}
From the above considerations, we 
conclude that the present approach is valid in 
the painted region (in yellow) in Fig. \ref{GL_param}. 
In the same figure, we have also shown the lines 
corresponding to the previously obtained results. 
The line indicated by \lq\lq Inaba {\it et al.}, 1996\rq\rq, 
is the line at which instability to C-AF order occurs\cite{Inaba:1996fm}. 
Within our theory, this corresponds to the line \lq\lq$A=0$\rq\rq. 
The line indicated by \lq\lq Yamase {\it et al.}, 2011\rq\rq, 
is obtained as the line at which $K(\vec{q})$ becomes negative for 
a certain $\vec{q}$\cite{Yamase:2011kf,spinon}. 

\subsection{Incommensurate Antiferromagnetic Order}

Let us examine the possible antiferromagnetic 
order based on $F^{(2)}$. 
In contrast to the ordinary GL free energy, the quadratic coefficient $B$ can be 
either positive or negative. 
In case of $B>0$, the free-energy minimum as a function of 
the wave vector $\vec{q}$ is located at the origin $\vec{q}=0$ and 
the corresponding order is C-AF. 
In case of $B<0$, the free-energy minima appear at  
non-zero $\vec{q}$ and incommensurate order becomes stable. 
In the latter case, we have two situations. 
Let us put $q_x = q \cos\theta$, $q_y = q \sin\theta$. 
Then $K(\vec{q})$ becomes 
\begin{align}
K(q,\theta) &= 
A + B q^2 + q^4 \left(C_1 + C'_2 \cos^2\theta \sin^2\theta\right) , 
\nonumber\\
&= 
A + Bq^2 + q^4 \left(C_1 + C'_2 \frac{1-\cos 4 \theta}{8}\right),
\end{align} 
where $C'_2 = C_2 - 2 C_1$. 

In case of $C'_2>0$, the minima of $K(q,\theta) $ 
as a function of $\theta$ lie at $\theta = \theta_n \equiv n\pi/2$ ($n$: integer) and 
\begin{align}
K\left(q,\theta_n\right) &= 
C_1\left(q^2 +\frac{B}{2 C_1}\right)^2 - \frac{B^2}{4 C_1} +A.
\end{align} 
$K(\vec{q})$ takes minimum at $q=q_{\rm min}\equiv\sqrt{|B|/(2 C_1)}$. 
%Note that $q_{\rm min}$ depends on $T$ and $\delta$. 
We call this order \lq\lq IC$_1$-AF\rq\rq. 

In case of $C'_2<0$, the minima of $K(q,\theta)$ 
lie at $\theta = \theta'_n \equiv (2n+1)\pi/4$ and 
\begin{align}
K\left(q,\theta'_n\right) &= \lambda
\left(q^2 +\frac{B}{2 \lambda}\right)^2 - \frac{B^2}{4 \lambda} +A
\end{align} 
where $\lambda = (C_1+C_2)/2$. 
$K(\vec{q})$ takes minimum at $q=q_{\rm min}\equiv\sqrt{|B|/(2 \lambda)}$. 
We call this order \lq\lq IC$_2$-AF\rq\rq. 

The positions of minima in $\vec{q}$-space are shown 
in the inset of Fig. \ref{FEIC-C} (color online). 

\subsection{The First Order Phase Transition}

In order to estimate the free energy including IC-AF, 
we restore the vector degrees of freedom of 
the order parameter, namely $M_j \rightarrow M_j \vec{n}_j\equiv \vec{M}_j$. 
Then the free energy can be written as 
\begin{align}
F = \sum_{\vec{q}}K(\vec{q}) |\vec{M}_{\vec{q}}|^2 
+\sum_j \left\{\beta_4 |\vec{M}_j|^4 + \beta_6 |\vec{M}_j|^6\right\}. 
\end{align}
We take the wave vector of the incommensurate order to be
parallel to $x$-axis, and 
put the order parameter as
\begin{align}
\vec{M}_j = (M \cos qx_j, M\sin qx_j,0),
\label{OPIC}
\end{align}
where $\vec{r}_j = (x_j,y_j)$. 
Note that if we take the order parameter to be 
unidirectional, such as $\vec{M}_j = (M \cos qx_j, 0,0)$, 
the amplitude $|\vec{M}_j|$ changes spatially, 
thus causing a free energy loss. 
Therefore, Eq. (\ref{OPIC}), whose magnitude is constant, 
may be most stable. 
The free energy corresponding to this order parameter is given by
\begin{align}
F = N\left\{K(q) M^2 
+\beta_4 M^4 + \beta_6 M^6\right\}.
\end{align}

Next, we minimize the free energy with respect to $M$. 
In case of $\beta_4>0$, we obtain the ordinary second order phase transition. 
However, in case of $\beta_4<0$, 
the first order phase transition occurs: 
\begin{enumerate}
\item[i)] In case of $K(q_{\rm min}) > |\beta_4|^2/(3 \beta_6)$, 
$F$ is a monotonic function of $M$ and always $M=0$ is stable. 
\item[ii)] In case of $|\beta_4|^2/(3 \beta_6) > K(q_{\rm min}) > |\beta_4|^2/(4 \beta_6)$, 
$F$ has metastable states, although $M=0$ is still stable. 
\item[iii)] In case of $|\beta_4|^2/(4 \beta_6)>K(q_{\rm min})>0$, 
non-zero $M$ becomes stable, although $M=0$ is still metastable. 
In the ordered state the free energy takes the value, 
\begin{align}
F=-\frac{\left(-\beta_4+\sqrt{\beta_4^2-3 k \beta_6}\right)^2\left(\beta_4+2\sqrt{\beta_4^2-3 k\beta_6}\right)}{27 \beta_6^2},
\label{FE1st}
\end{align}
where $k=K(q_{\rm min})$. 
\item[iv)] In case of $0>K(q_{\rm min})$, $M=0$ becomes unstable and 
only the ordered state is stable. 
\end{enumerate}
As seen from the above, the system undergoes the first order phase transition at 
the temperature defined by $|\beta_4|^2/(4 \beta_6) = K(q_{\rm min})$.

\section{Mean Field Phase Diagram}

\begin{figure}[h]
\begin{centering}
\includegraphics[clip,width=8cm]{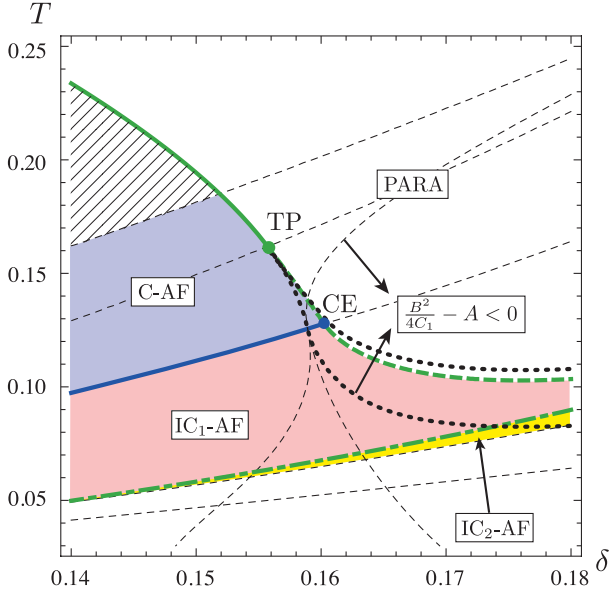} 
\par\end{centering}
\caption{(Color online) Phase diagram obtained by the present analysis. 
Bold solid (green) line shows the second order phase transition and bold dashed line (green) 
shows that of first order. 
PARA is the paramagnetic phase. 
IC$_1$-AF and IC$_2$-AF are incommensurate antiferromagnetic phase, 
with different incommensurate wave vectors (see Fig.\ref{FEIC-C} for detail), 
and bold dashed-dotted line (green) is the boundary between them. 
Thin broken lines are the reproduction of those shown in Fig. \ref{GL_param}. 
The area between two dotted lines (black) is the 
region where metastable states associated with the first order 
phase transition exist. 
TP is the tricritical point and CE is the critical endpoint. }
\label{phase_diag} %
\end{figure}

Now we are ready to discuss the mean-field phase diagram.  
From the arguments given above, we found 
several phases shown in Fig. \ref{phase_diag} (color online). 

In contrast to the preceding results\cite{Yamase:2011kf}, 
where the existence of IC-AF is pointed out between 
\lq\lq Inaba {\it et al.}\rq\rq -line and \lq\lq Yamase {\it et al.}\rq\rq-line 
in Fig. \ref{GL_param}, we found a significantly large region of IC-AF in Fig. \ref{phase_diag}. 
(IC-AF region is divided into IC$_1$-AF and IC$_2$-AF. 
However we do not consider the result of the IC$_2$-AF is 
completely reliable, since GL expansion is not good in this region. 
Therefore we do not take this phase seriously in this paper. )
Though the present treatment is not applicable in the hatched region, 
C-AF phase may be occupying this region also. 

Next we discuss the nature of the phase transitions. 
We found a first order transition line, shown by the 
bold dashed line (green) in Fig. \ref{phase_diag}. 
This line is defined by $K(q_{\rm min}) = |\beta_4|^2/(4 \beta_6)$, 
which corresponds to the temperature 
at which the free energy of Eq. (\ref{FE1st}) becomes 
negative, namely smaller than the free energy of the paramagnetic state $M=0$. 
This first-order-transition line extends up to $\beta_4=0$ line 
(see Fig. \ref{GL_param}) and 
is connected to the second-order-transition line in the region of $\beta_4>0$.  
This connecting point, indicated by TP in Fig. \ref{phase_diag}, is called the tricritical point 
\cite{chaikin2000principles}.  
Two dotted lines (black) are defined by 
$|\beta_4|^2/(3 \beta_6) = K(q_{\rm min})$ (upper line) and 
$B^2/(4 C_1) -A =0$ (lower line). 
In the region between these two lines, metastable states exist. 
We can estimate the strength of the first order 
phase transition, following Ref. \citen{Halperin:1974vx}, 
by the width of the temperature region, 
where the metastable states exist. 
In our case, the first-order nature is stronger in the 
PARA to IC$_1$-AF transition and 
weaker in the PARA to C-AF transition. 
The lower dotted line shows the temperature below which 
$K(\vec{q})$ becomes negative at a certain $\vec{q}$. 
Therefore this line corresponds to the phase transition line 
obtained by Yamase {\it et al.}\cite{Yamase:2011kf} (shown in Fig. \ref{GL_param}). 
Comparing these lines, we can see that GL expansion is overestimating 
the transition temperature as compared to 
the study by Yamase {\it et al.}, obtained by treating exact 
$\vec{q}$-dependence of $K(\vec{q})$, 
especially at higher doping region ($\delta > 0.16$).

\begin{figure}[h]
\begin{centering}
\includegraphics[clip,width=8cm]{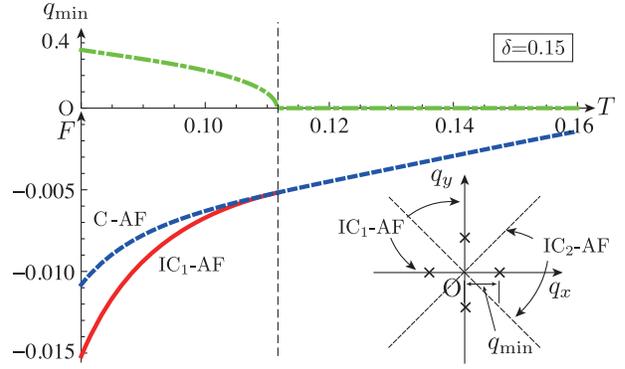} 
\par\end{centering}
\caption{(Color online) Plot of the free energy of C-AF and IC-AF states 
as a function of temperature at $\delta=0.15$. 
The development of the incommensurate wave number $q_{\rm min}$ 
in the IC-AF phase is also shown in the upper graph.
The inset shows the locations of incommensurate wave vectors 
in IC$_1$-AF state. 
In case of IC$_2$-AF state the positions are rotated by 45$^\circ$ around 
the origin. }
\label{FEIC-C} %
\end{figure}

Finally we study the phase transition between C-AF and IC-AF. 
In Fig. \ref{FEIC-C} we have plotted the free energy of 
C-AF state ($\vec{q}=0$) and IC-AF state ($\vec{q}=(q_{\rm min},0)$) 
along the line $\delta = 0.15$. 
The wave number $q_{\rm min}$ is also shown in a same plane. 
The solid curve (IC$_1$-AF, in red)  becomes lower 
than dashed curve (C-AF, in blue)
at the temperature below $T \sim 0.112$, and 
the connection between two curves seems to 
be smooth. 
This probably suggests the second order phase transition 
between C-AF and IC-AF phase. 
Therefore the intersection of the C-IC boundary and 
the first-order-transition line, namely CE in Fig. \ref{phase_diag}, 
is the so-called critical endpoint, at which the second-order-transition line 
ends at the first-order-transition line\cite{chaikin2000principles}.  

\section{Discussion}

In this paper we have clarified the mean-field magnetic phase diagram of the 
$t$-$J$ model within the GL approximation. 
Obtained phase diagram contains a significantly large area 
of IC-AF phase, as compared to the previous theories\cite{Inaba:1996fm,Yamase:2011kf}. 
At this stage, these findings may not have direct relationships to the 
experimentally obtained phase diagram\cite{JULIEN:2003ec,Mukuda:2011km}. 
However they include 
several clues to the quantitative understanding of the 
high-$T_c$ oxides based on $t$-$J$ model. 

Firstly, our results provide an insight into the 
effects of the thermal fluctuations in high-$T_c$ oxides. 
In the conventional GL free energy, 
the quadratic coefficient as a function of wave number is 
assumed to be constant near the critical points. 
In our free energy, $\vec{q}$-dependence of $K(\vec{q})$ 
is more complicated. 
Especially, at the boundary between C-AF and IC-AF, 
the quadratic coefficient $B$ vanishes, 
which may cause a significant enhancement of the 
fluctuation effects. 
There is a possibility that C-AF area 
is strongly reduced in the phase diagram 
if we take into account the thermal fluctuations. 
Further research is ongoing in this direction.  
It may be also interesting to pursue possibilities of 
understanding pseudogap phase in relation to 
antiferromagnetic fluctuations, 
although in this case RVB order cannot probably be neglected.  

Another important issue may be the 
existence of widely spread IC-AF phase. 
We may consider that this phase has some relationship 
with the experimentally observed \lq\lq spin glass\rq\rq 
in high-$T_c$ oxides\cite{Mukuda:2011km}. 
The most important difference between C-AF and IC-AF phase is 
that multiple stable states, 
namely $\vec{q}=(\pm q_{\rm min},0)$ or $(0, \pm q_{\rm min})$, 
exist only in the latter phase (see the inset of Fig. \ref{FEIC-C}). 
In the presence of randomness, there can exist 
domains of IC-AF orders with different $\vec{q}$\,'s 
and such a behavior is similar to spin glass system. 
Actually our GL free energy is similar to that 
discussed by Sherrington\cite{Sherrington:1980wp}
in relation to the spin glass. 
According to that literature, 
the non-trivial $\vec{q}$-dependence 
(or the frustration) is crucial for the 
appearance of the spin glass order\cite{Sherrington:1980wp,Ma:1978wm}. 
It is also interesting that Yamase {\it et al.} has pointed out the relation between 
IC-AF to charge stripes\cite{Yamase:1999fm}. 
Although it is just a possibility at this stage, the relation of IC-AF phase to 
experimentally observed spin glass phase and the charge stripes 
may be worth further studies.

\section{Summary}

We have examined the antiferromagnetic phase diagram of 
high-$T_c$ oxides based on $t$-$J$ model and GL expansion. 
It has been found that the incommensurate antiferromagnetic phase 
occupy larger area in the phase diagram than 
the preceding works\cite{Inaba:1996fm,Yamase:2011kf}, 
and the transition from paramagnetic 
to antiferromagnetic (commensurate or incommensurate) 
phase can be first order depending on the doping. 
Possibilities of tricritical point and critical endpoint 
in the magnetic phase diagram are pointed out.  
Existence of the incommensurate antiferromagnetic 
phase may enhance the 
effects of thermal fluctuations and it may also 
lead to glass-like behavior in the presence of 
randomness. 

\begin{acknowledgment}

%\acknowledgment
This work was supported by JSPS KAKENHI Grant Numbers 24540392, 22540329. 

\end{acknowledgment}

\appendix
\section{GL coefficients: $\vec{q}$-dependences}
\label{GLcoeff1}

We calculate the quadratic GL coefficient $K(\vec{q})$ 
to the fourth order in the wave vector $\vec{q}$. 
This process can be carried out by applying Taylor series 
expansion with respect to $\vec{q}$ and performing the 
$\vec{k}$-summation over the Brillouin zone
($|k_x|,|k_y|<\pi)$ in Eq. (\ref{f2}). 
Basically this summation (integral) is not singular, however, 
if we expand it with respect to the wave numbers, $q_x$ and $q_y$, 
removable singularities (such as \lq\lq 0/0\rq\rq) appear on the lines $|k_x|+|k_y| = \pi$. 
In the numerical calculation, this type of singularity also causes errors. 
To avoid this, it is advantageous to introduce the new coordinates $(k_s,k_t)$ by
$k_x = (k_s+k_t)/\sqrt{2}$, $k_y = (k_s-k_t)/\sqrt{2}$. 
Then the lines $|k_x|+|k_y| = \pi$ become parallel (or perpendicular) to the axes. 
In this coordinate, $\xi_{\vec{k}} = -4 \tilde{t} \cos (k_s/\sqrt{2}) \cos (k_t/\sqrt{2}) - \mu$, 
and the vector $\vec{Q}$ becomes $(0,\sqrt{2}\pi)$. 
We further introduce 
\begin{align}
\eta_{\vec{k}} &= 8 \tilde{t} \cos (k_s/\sqrt{2}) \cos (k_t/\sqrt{2}),
\nonumber\\
\delta \xi_{\vec{k},\vec{q}} &= 
\xi_{\vec{k}+\vec{Q}+\vec{q}}-\xi_{\vec{k}+\vec{Q}}.
\end{align}
Then we obtain
\begin{align}
\xi_{\vec{k}+\vec{Q}} &= 
\xi_{\vec{k}}+\eta_{\vec{k}},
\nonumber\\
\xi_{\vec{k}+\vec{Q}+\vec{q}} -\xi_{\vec{k}} &=
\xi_{\vec{k}+\vec{Q}+\vec{q}}-\xi_{\vec{k}+\vec{Q}} 
+(\xi_{\vec{k}+\vec{Q}} -\xi_{\vec{k}})
\nonumber\\
&=\delta \xi_{\vec{k},\vec{q}}+\eta_{\vec{k}},
\end{align}
and 
\begin{align}
\chi^{(0)} (\vec{q}) = \frac{1}{N}\sum_{\vec{k}\in{\rm BZ}}
\frac{\tanh(\beta (\xi+\eta+\delta \xi)/2)
- \tanh(\beta\xi/2)}{\eta + \delta \xi},
\label{chi0-1}
\end{align}
where we have suppressed the arguments $\vec{k}$ and $\vec{q}$. 
Since all the $\vec{q}$ dependences are contained in $\delta\xi$, 
we first expand Eq. (\ref{chi0-1}) with respect to $\delta\xi$ and, then, 
expand $\delta\xi$ with respect to $\vec{q}$. 
Note that, because $\delta\xi$ is the first order or higher in $q_s$ or $q_t$, 
the expansion up to the fourth order in $\delta\xi$ is 
sufficient for our purpose. 
Let us write
\begin{align}
\chi^{(0)}(\vec{q}) =& \frac{1}{2N}\sum_{\vec{k}\in {\rm BZ}}
\sum_{n=0}^{4} \alpha^{(n)}(\vec{k}) (\delta\xi)^n,
\\
(\delta\xi)^n =& \nu_a^{(n)}(\vec{k})+\nu_b^{(n)}(\vec{k}) (q_s^2+q_t^2)
\nonumber\\
&\phantom{aaa}+\nu_{c1}^{(n)}(\vec{k}) (q_s^4+q_t^4)
+\nu_{c2}^{(n)}(\vec{k}) q_s^2 q_t^2,
\end{align}
where $\alpha^{(n)}(\vec{k})$ is the expansion coefficient 
of the $n$-th order, 
then $\chi^{(0)}(\vec{q})$ reads,
\begin{align}
\chi^{(0)}(\vec{q}) = c_a + c_b (q_s^2+q_t^2) + 
c_{c1}(q_s^4+q_t^4) + c_{c2} q_s^2 q_t^2
\label{chi0-st}
\end{align}
with 
\begin{align}
c_a =& \frac{1}{8\pi^2} \frac{1}{N}\sum_{\vec{k}\in {\rm BZ}}
\sum_{n=0}^{4} \alpha^{(n)}(\vec{k}) \nu_a^{(n)}(\vec{k}) 
\nonumber\\
=&\frac{1}{8\pi^2}  \frac{1}{N}\sum_{\vec{k}\in {\rm BZ}} \,
\vec{\alpha}(\vec{k})\cdot\vec{\nu_a}(\vec{k}),\,\,\,
\cdots, {\rm etc.},
\end{align}
where $\vec{\alpha}$ and $\vec{\nu}_a$ are five dimensional vector
whose $n$-th component being $\alpha^{(n)}$ and $\nu^{(n)}_a$, respectively. 
By using the abbreviations, 
\begin{align}
C_s &= \cos\left({{k_s}}/{\sqrt{2}}\right), 
C_t = \cos\left({{k_t}}/{\sqrt{2}}\right), 
\nonumber\\
S_s &= \sin\left({{k_s}}/{\sqrt{2}}\right), 
S_t = \sin\left({{k_t}}/{\sqrt{2}}\right),
\nonumber\\
C'_s &= \cos\left({\sqrt{2}}{{k_s}}\right), 
C'_t = \cos\left({\sqrt{2}}{{k_t}}\right), 
\nonumber\\
S'_s &= \sin\left({\sqrt{2}}{{k_s}}\right), 
S'_t = \sin\left({\sqrt{2}}{{k_t}}\right),
\end{align}
the vector $\vec{\nu}_a,\,\cdots$, etc. are given by
\begin{align}
&\vec{\nu}_a = (\nu_a^{(0)},\cdots, \nu_a^{(4)}) = 
(1,0,0,0,0),
 \\
 &\vec{\nu}_b =  \left(0,-{\tilde{t}}{C_s} {C_t} ,-2
   {\tilde{t}}^2 ({C'_s}
   {C'_t}-1),0,0\right),
   \\
   & \vec{\nu}_{c1} = \biggl(0,\frac{{\tilde{t}}{C_s} {C_t}
   }{24},\frac{1}{12} {\tilde{t}}^2
   \left\{{C'_s} (7 {C'_t}+3)+3
   {C'_t}-1\right\},
      \nonumber\\
   &\phantom{aaaa}
   6 {\tilde{t}}^3{C_s} {C_t}
    ({C'_s} {C'_t}-1),
   32
   {\tilde{t}}^4 \left({C_s}^4
   {S_t}^4+{C_t}^4
   {S_s}^4\right)\biggr),
   \\
   & \vec{\nu}_{c2} = \biggl(
0,\frac{ {\tilde{t}}{C_s} {C_t}
  }{4},\frac{1}{2} \tilde{t}^2
   \left\{{C'_s} (7 {C'_t}-1)-{C'_t}-1\right\},
       \nonumber\\
   &\phantom{aaa}
   12 {\tilde{t}}^3
   {C_s} {C_t} \left\{{C'_s}
   (3 {C'_t}-2)-2 {C'_t}+1\right\},24\tilde{t}^4
   {S'_s}^2 {S'_t}^2
   \biggr).
 \end{align}
 The vector $\vec{\alpha}$ is given, using 
$\tau_{\pm} =  \tanh \left\{(\eta \pm2 \mu )/(4
   {T})\right\}
$, as
 \begin{align}
\vec{\alpha} = \biggl(&
 \frac{{\tau_{+} }+{\tau_{-} }}{8 \pi ^2 \eta}
 ,
 \frac{-\eta  {\tau_{-} }^2+\eta -2
   {T} ({\tau_{+} }+{\tau_{-} })}{16 \pi ^2
   \eta ^2 {T}},
   \nonumber\\
&
   \frac{\eta ^2 {\tau_{-} }
   \left({\tau_{-} }^2-1\right)+2 \eta 
   {T} \left({\tau_{-} }^2-1\right)+4
   T^2 ({\tau_{+} }+{\tau_{-} })}{32 \pi
   ^2 \eta ^3 T^2}
   ,
  \nonumber\\
&
   \frac{1}{192 \pi ^2 \eta ^4 T^3}
   \biggl\{\eta ^3
   \left(-3 {\tau_{-} }^4+4
   {\tau_{-} }^2-1\right)
   \nonumber\\
  &\phantom{aaaaa} -6 \eta ^2 {T}
   {\tau_{-} } \left({\tau_{-} }^2-1\right)
   -12
   \eta  T^2
   \left({\tau_{-} }^2-1\right)
   \nonumber\\
   &\phantom{aaaaaaaaaaa}
   -24 T^3
   ({\tau_{+} }+{\tau_{-} })\biggr\}
   ,
  \nonumber\\
&
   \frac{1}{384 \pi ^2 \eta ^5 T^4}
   \biggl\{
   \eta ^4 {\tau_{-} }
   \left(3 {\tau_{-} }^4-5
   {\tau_{-} }^2+2\right)
      \nonumber\\
   &\phantom{aaaaa}
+2 \eta ^3 {T}
   \left(3 {\tau_{-} }^4-4
   {\tau_{-} }^2+1\right)
  \nonumber\\
   &\phantom{aaaaa}
 +12 \eta ^2
   T^2 {\tau_{-} }
   \left({\tau_{-} }^2-1\right)
   +24 \eta 
   T^3 \left({\tau_{-} }^2-1\right)
     \nonumber\\
   &\phantom{aaaaa}
+48
   T^4 ({\tau_{+} }+{\tau_{-} })
     \biggr\}
  \biggr).
  \label{alpha}
   \end{align}
 Since $\eta$ vanishes on the lines $|k_x|+|k_y| = \pi$ ($k_s, k_t =\pm \pi/\sqrt{2}$), 
 $\vec{\alpha}$ seems to diverge on these lines due to 
the powers of $\eta$ in the denominators. 
This is just superficial, because the numerators also vanish on the same 
lines. 
To avoid this singularity, we use Taylor series expansion 
with respect to $k_s$ and $k_t$ in the narrow region near 
the lines $|k_x|+|k_y| = \pi$, 
instead of using the original expression, Eq. (\ref{alpha}). 

From Eqs.(\ref{f2}) and (\ref{chi0-st}), the GL coefficients in Eq. (\ref{k-def})  
are given by 
\begin{align}
A &=  J (1-2 J c_a),\\
B &= -J^2\left(\frac{c_a}{2} - 2 c_b\right),\\
C_1 &= J^2 \left(\frac{c_a}{24}-\frac{c_b}{2}+2c_{c1}\right),\\
C_2 &= -J^2\left(-c_{b}+2 c_{c2}\right).
\end{align}

\section{GL coefficients: higher order in $M_{j}$}
\label{GLcoeff2}

If the order parameter $M_j$ is a constant $M$, 
the Hamiltonian, Eqs. (\ref{Hamiltonian}) and (\ref{MFH}), is rewritten by 
limiting the summation with respect to $\vec{k}$ 
from the original Brillouin zone ($|k_x|$,$|k_y|<\pi$) to 
the magnetic Brillouin zone (MBZ) ($|k_x|+|k_y|<\pi$) as
\begin{align}
H =& \sum_{\sigma, \vec{k}\in {\rm MBZ}}
\left(\begin{array}{c c}
f_{\vec{k}\,\sigma}^\dagger&
f_{\vec{k}-\vec{Q}\,\sigma}^\dagger
\end{array}
\right)\cdot
\nonumber\\
&
\left(
\begin{array}{cc}
\xi_{\vec{k}} & -i J (-1)^\sigma M\\
-i J (-1)^\sigma M & \xi_{\vec{k}-\vec{Q}}
\end{array}
\right)\cdot
\left(
\begin{array}{l}
f_{\vec{k}\,\sigma}\\
f_{\vec{k}-\vec{Q}\,\sigma}
\end{array}
\right)+2 J M^2.
\end{align}
Since $\xi_{\vec{k}}=-2\tilde{t}\gamma_{\vec{k}}-\mu$ and 
$\xi_{\vec{k}-\vec{Q}}=2\tilde{t}\gamma_{\vec{k}}-\mu$, 
we obtain the quasiparticle energy as 
\begin{align}
E = \pm\sqrt{(2 \tilde{t}\gamma_{\vec{k}})^2 + (2 J M)^2}-\mu
\equiv\pm E_{\vec{k}}-\mu.
\end{align}
Then the partition function is obtained following the 
ordinary recipe as, 
\begin{align}
Z = \prod_{\vec{k}\in {\rm MBZ}}
\left\{(1+e^{-(E_{\vec{k}}-\mu)/T})(1+e^{(E_{\vec{k}}+\mu)/T})\right\}^2e^{- 2 J M^2/T}
\end{align}
and the free energy (or potential of $M$) is 
obtained by $U(M) = -T \ln Z$. 
The same formulae are given in Ref \citen{Yamase:2011kf}. 

\vfill\eject
%\bibliographystyle{jpsj}
%\bibliography{hightc}

\begin{thebibliography}{10}
\bibitem{Kq}
Exactly speaking, this is not necessarily the condition for $C_1$ and $C_2$, since 
the integration with respect to $\vec{q}$ is limited to a finite area ($|q_x|$, $|q_y|<\pi$) 
and no divergence results even if some of $C_1$, $C_2$, $\cdots$ etc. are negative. 
However, we put this condition just for the rough standard of the ultraviolet fidelity of our model.
\bibitem{spinon}
Note that in Refs. \citen{Inaba:1996fm} and \citen{Yamase:2011kf}, 
not only the AF order but also the 
RVB order is considered, although we have neglected the latter in this paper. 
As far as the AF order is concerned, this does not seem to cause 
significant differences in the phase diagram.
\end{thebibliography}

\end{document}